\documentstyle[epsf,amssymb]{article}

\begin{document}
\begin{center}

{\Large{\bf Phase Structure and Critical Behavior of
Multi-Higgs U(1) Lattice Gauge Theory in Three Dimensions}}\\
\vspace{1cm}
{\Large Tomoyoshi Ono$^1$, Shunsuke Doi$^2$, Yuki Hori$^2$,
Ikuo Ichinose$^1$ and Tetsuo Matsui$^2$} \\
{ $^1$Department of Applied Physics, Graduate School of Engineering, \\
Nagoya Institute of Technology, 
Nagoya, 466-8555 Japan} \\
{ $^2$Department of Physics, Kinki University, 
Higashi-Osaka, 577-8502 Japan}

{\bf abstract}
\end{center}
We study the three-dimensional (3D) compact U(1) lattice gauge theory
coupled with $N$-flavor Higgs fields by means of the Monte Carlo simulations. 
This model is relevant to
multi-component superconductors, antiferromagnetic
spin systems in easy plane, inflational cosmology, etc. 
It is known that there is no phase transition in the $N=1$ model.
For $N=2$, we found that 
the system has a second-order phase transition line $\tilde{c}_1(c_2)$
in the $c_2$(gauge coupling)$-c_1$(Higgs coupling) plane, 
which separates the confinement phase and the Higgs phase.  
Numerical results suggest that the phase transition belongs to 
the universality class of the 3D XY model as the previous works
by Babaev et al. and Smiseth et al. suggested.
For $N=3$, we found that there exists a critical line similar to that in 
the $N=2$ model, but the critical line is separated into two parts;
one for $c_2 < c_{2{\rm tc}}=2.4\pm 0.1$ with first-order transitions, 
and the other for
$ c_{2{\rm tc}} < c_2$ with second-order transitions, indicating the 
existence of
a tricritical point. We verified that similar phase diagram 
appears for the $N=4$ and $N=5$ systems.
We also studied the case of anistropic Higgs coupling in the $N=3$ model
and found that there appear two second-order phase transitions or
a single second-order transition and a crossover depending on the
values of the anisotropic Higgs couplings.
This result indicates that an ``enhancement" of phase transition occurs
when multiple phase transitions coincide at a certain point in 
the parameter space.

\newpage

\section{Introduction}

There are many interesting physical systems involving multi-component
($N$-component) matter fields. 
Sometimes they are associated
with exact or approximate symmetries like ``flavor" symmetry.  
In some cases, the large-$N$ analysis or the $1/N$ 
expansion\cite{largen} for an $N$-flavor system 
is applicable and it gives us useful 
information that cannot be obtained by the ordinary perturbative 
calculations. 
But the properties of the large-$N$ systems may differ
from those at medium values of $N$ that one actually wants to know. 
Study of the $N$-dependence of various systems is certainly 
interesting, but has not been examined well. 

Among these ``flavor" physics, the effect of matter 
fields upon gauge dynamics is of quite general
interest in quantum chromodynamics, strongly correlated 
electron systems, quantum spins, etc.\cite{gauge1,gauge2,gauge3,gauge4}. 
In the present paper,  
we shall study the three-dimensional (3D) U(1) gauge theory
with multi-component Higgs fields $\phi_a(x)
\equiv |\phi_a(x)| \exp(i\varphi_a(x))\
 (a=1,\cdots,N)$ with fixed amplitudes $|\phi_a(x)|=1$.
This model is of general interest, and knowledge of 
its phase structure, order of its phase 
transitions, etc. may be useful to get better 
understanding of various physical systems.
These systems include the following:

{\it $N$-component superconductor:}
Babaev\cite{babaev} argued  that under a high pressure and 
at low temperatures hydrogen gas may become a liquid and
exhibits a transition from a superfluid to a superconductor.
There are two order parameters; $\phi_e$ for electron pairs
and $\phi_p$ for proton pairs. They may be treated as 
two complex Higgs fields ($N=2$). In the superconducting phase, 
both $\phi_e$ and $\phi_p$ develop an off-diagonal long-range 
order, while in the superfluid phase, only the 
neutral order survives; $\lim_{|x| \rightarrow \infty} \langle
\phi_e(x) \phi_p(0)\rangle \neq 0$. 
 
{\it $p$-wave 
superconductivity of cold Fermi gas:}
Each fermion pair in a $p$-wave superconductor
has angular momentum $J=1$ and the order parameter 
has three components, $J_z= -1,0,1$. They are regarded
as three Higgs fields ($N=3$). As the strength of attractive
force between fermions is increased,  a crossover from
a superconductor of the BCS type  to the type of  
Bose-Einstein condensation is expected to take place\cite{ohashi}.

{\it Phase transition of 2D antiferromagnetic(AF) 
spin models:}
In the $s=1/2$ AF spin models, a phase transition occurs from the N\'eel 
state to the valence-bond solid state as parameters are varied. 
Senthil et al.\cite{senthil}
argued that the effective theory describing
this transition take a form of  
 U(1) gauge theory of spinon ($CP^1$) field $z_a(x)\
(|z_1|^2+|z_2|^2=1)$ with an additional Berry phase 
and a ``deconfined phase" of spinons appears at the
critical point.
If one consider the easy-plane limit ($S_z=0$), $|z_1|^2=|z_2|^2=1/2$
and the $CP^1$ bosons are expressed by two Higgs fields as
$z_a= \exp(i\varphi_a)/\sqrt{2}$ ($N=2$).
This system is studied by Nogueira et al. by a renormalization
group analysis\cite{RG} and it is clarified that the above
easy-plane limit has only a first-order phase transition.

Similar limit to the above may be taken to relate
the superconductivity of ultracold fermionic atoms  
with spin $J$ to the U(1) gauge model with $N$ Higgs fields. 
J. Zhao, et al. considered the $SU(N)$ Hubbard model
to describe the superconductivity of fermionic atoms,
which has a $N=2J+1$-component order parameter\cite{cpn-1}. 
At large repulsion $U$ and at the filling factor $n=1/N$,
the model becomes the U(1) gauge model with $CP^{N-1}$ spins. 
A $CP^{N-1}$ variable
is parametrized as $z_a=\rho_a\exp(i\varphi_a)$
with $\sum_{a=1}^N \rho_a^2=1$. In the symmetric limit,
which is the easy-plane limit for $N=2$,
$\rho_a^2= 1/N$ and $z_a$ becomes a Higgs field.

Effects of doped fermionic holes (holons) 
to these AF spins were also studied extensively. 
The effective theory obtained by integrating out holon 
variables may be a U(1) gauge theory with $N=2$ Higgs fields
(with nonlocal gauge interactions).
Kaul et al.\cite{kaul} predicts that such a system exhibits
a second-order transition, while numerical 
simulations of Kuklov et al.\cite{kuklov} exhibit 
a weak first-order transition.
This point should be clarified in future study.

{\it Inflational cosmology:}
In the inflational cosmology\cite{guth}, a set of Higgs 
fields is introduced to describe a phase transition
and inflation in early universe.
Plural Higgs fields are necessary
in a realistic model\cite{allahverdi}.

In the rest of the present paper, we shall study the multi-Higgs 
lattice models
by Monte Carlo(MC) simulations. 
We consider the simplest form of the model, i.e., the
3D {\em compact} lattice gauge theory without Berry's phase.
We introduce  the Higgs fields $\phi_{xa}$ on the site $x$ 
of the cubic lattice and treat them in the London limit, 
$|\phi_{xa}|=1,\ \phi_{xa}=\exp(i\varphi_{xa})$.  
We also put the compact U(1) 
gauge field $U_{x\mu} = \exp(i\theta_{x\mu})$
on the link $(x,x+\mu)$. $\mu (=1,2,3)$ is the direction 
index (we use them also as the unit vectors). 
The action $S$ consists of the Higgs coupling
with its coefficients $c_{1a}\; (a=1,\dots, N)$ 
and the plaquette term with its coefficient $c_2$ as 
\begin{equation}
S={1\over 2}\sum_{x,\mu}\sum_{a=1}^N
\Big(c_{1a}\phi^\dagger_{x+\mu,a}
U_{x\mu}\phi_{xa}+\mbox{H.c.}\Big) 
+{c_2 \over 2}\sum_{x,\mu<\nu}
(U^\dagger_{x\nu}U^\dagger_{x+\nu,\mu}U_{x+\mu,\nu}U_{x\mu}
+\mbox{H.c}).
\label{action}
\end{equation}
The partition function $Z$
of the model is given by
\begin{eqnarray}
Z &=& \int[dU][d\phi] \exp(S),\nonumber\\
\int[dU][d\phi]&=& \prod_{x,\mu}\int_{-\pi}^{\pi}
\frac{d\theta_{x\mu}}{2\pi}
\prod_{x,a}\int_{-\pi}^{\pi}\frac{d\varphi_{xa}}{2\pi}.
\label{z}
\end{eqnarray}

A couple of models close to Eqs.(\ref{action}) and (\ref{z}) 
have been investigated.
Smiseth et al.\cite{smiseth} studied the {\em noncompact} U(1) 
Higgs models. A duality transformation maps the charged 
sector into the inverted $XY$ spin model. 
Thus they predicted that the system exhibits a single 
inverted $XY$ transition and $N-1$ $XY$ transitions. 
Their numerical study confirmed this prediction for $N=2$.
For $N=2$, Kragset et al.\cite{kragset} studied the effect
of Berry's phase term in the compact Higgs model. 
They reported that Berry's phase term suppresses monopoles
(instantons) and changes the second-order phase transitions
to first-order ones.

The phase structure of the present system (\ref{action}) 
can be studied by the following consideration developed 
by Smiseth et al.\cite{smiseth,smiseth2}.
That is,
among $N$ phases $\varphi_a(x)$ of the Higgs fields,
the sum $\tilde{\varphi}_{+}\equiv\sum_a\varphi_a$ couples
to the gauge field and describes charged excitations, 
whereas the remaining $N-1$ independent linear combinations
$\tilde{\varphi}_{i} (i=1,\cdots,N-1)$ 
describe neutral excitations. The latter $N-1$ modes
may be regarded as a set of $N-1$ $XY$ spin models.
As the $N=1$ compact U(1) Higgs
model stays always in the confinement phase\cite{janke}, 
we expect $N-1$ second-order transitions of the type of 
the $XY$ model. 
The above discussion is useful to get an 
intuitive picture of the phase structure.
In the present paper,
we shall study the system (\ref{action}) by means 
of MC simulation and verify the above conclusion.

The rest of the present paper is organized as follows.
In Sec.2, we shall study the $N=2$ multi-Higgs model and report 
the results of the MC simulations, in particular, its phase diagram
in the $c_1-c_2$ plane.
We measure the internal energy, specific heat and instanton density.
In Sec.3, we shall investigate the $N=3$ cases. 
By varying the ratios of the three Higgs couplings, the model exhibits
interesting phase structure. In Sec.4, we present the result
of the cases of $N=4$ and $N=5$.
Section 5 is devoted for conclusion.

\section{Two-flavor Higgs model ($N=2$)}
\setcounter{equation}{0}

For the MC simulations, we used the standard Metropolis
algorithm\cite{metropolis}.
We consider the cubic lattice with the  
periodic boundary condition and its size $L^3$ up to $L = 36$. 
The typical statistics used was $6\cdot 10^5$ MC steps 
and the averages and errors were estimated over 20 samples.

We first study the $N=2$ case with symmetric couplings 
$c_{11}=c_{12} \equiv c_1$.
We measured the internal energy $U\equiv -\langle S \rangle/L^3$
and the specific heat $C\equiv\langle 
(S-\langle S\rangle)^2\rangle/L^3$ in 
order to obtain the phase diagram and determine the order of 
phase transitions.

\begin{figure}[htbp]
\begin{center}
\epsfxsize=7cm
\epsffile{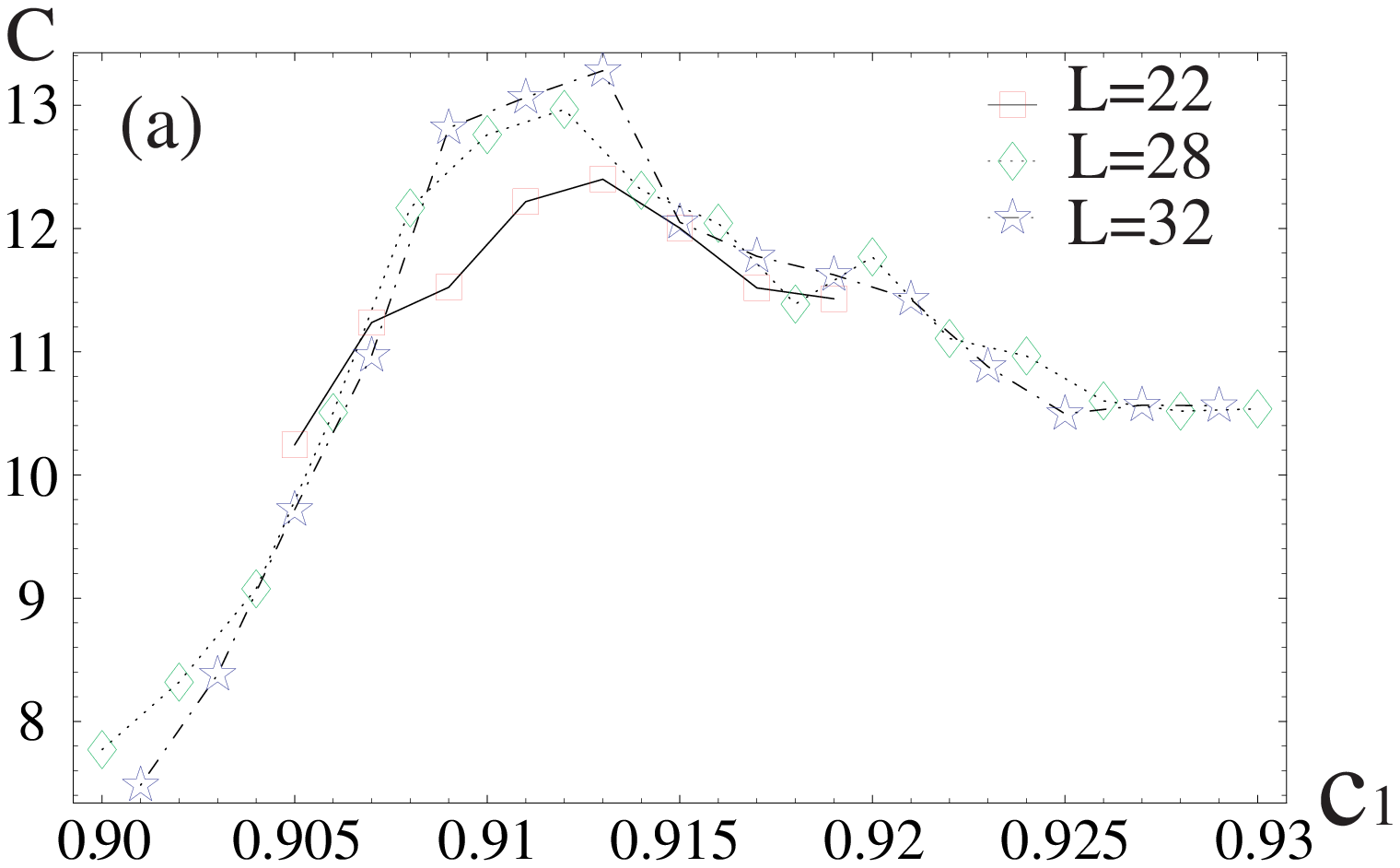}
\epsfxsize=6.7cm
\epsffile{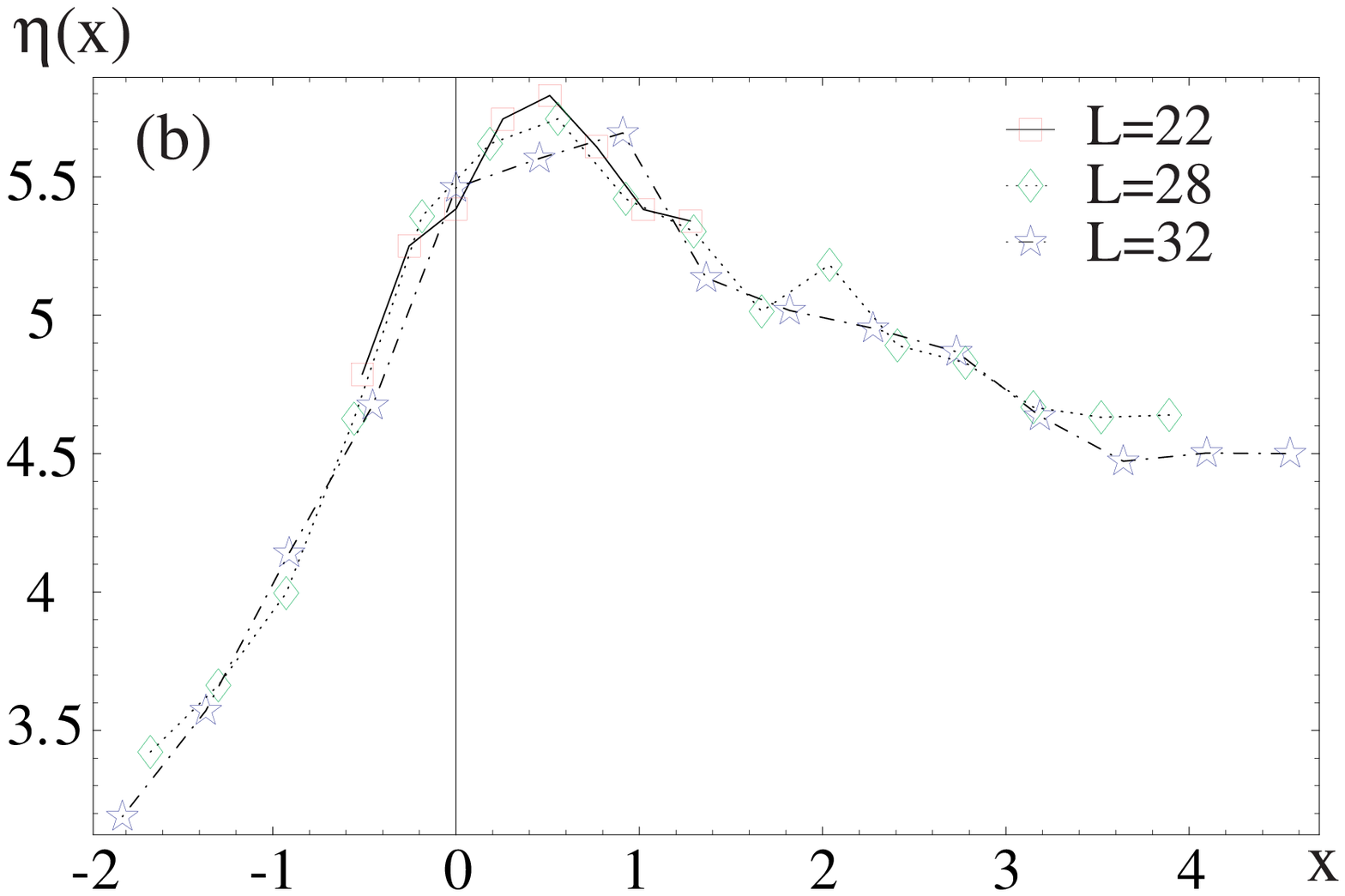}
\caption{(a) System-size dependence of specific heat $C$ of
$L=22,28,32$
for $N=2$ along $c_2=0.4$.
(b) Scaling function $\eta(x)$ of Eq.(\ref{FSS}) for Fig.\ref{fig1}(a).
}
\label{fig1}
\end{center}
\end{figure}

In Fig.\ref{fig1}(a), we show $C$ at $c_2=0.4$ as a function of $c_1$
for $L=22,28,32$.
The peak of $C$ develops as the system size is increased. 
The results indicate that a second-order phase transition occurs at 
$c_1 \simeq 0.91$.
In fact, we applied the finite-size-scaling (FSS) hypothesis 
to $C$ of Fig.\ref{fig1}(a) in the form of 
\begin{equation}
C(c_1,L)=L^{\sigma/\nu}\eta(L^{1/\nu}\epsilon), 
\label{FSS}
\end{equation}
where 
$\epsilon=(c_1-c_{1\infty})/c_{1\infty}$ and $c_{1\infty}$ 
is the critical coupling at $L\rightarrow \infty$.
We determined $\nu=0.65\pm 0.02, \; \sigma=0.16\pm 0.02$, and 
$c_{1\infty}=0.909\pm 0.010$ with 
 the scaling function $\eta(x)$ 
plotted in Fig.\ref{fig1}(b).
This result supports the FSS (\ref{FSS}). 

The above results for $N=2$ are consistent with the 
prediction given in the introduction. 
The sum $\tilde{\varphi}_{x+}\equiv \varphi_{x1}+\varphi_{x2}$
couples with the compact gauge field and generates no
phase transition\cite{janke}, while
the difference $\tilde{\varphi}_{x-}\equiv 
\varphi_{x1}-\varphi_{x2}$ 
behaves like the angle variable in the 3D $XY$ model.
The 3D $XY$ model has a second-order phase transition 
with the critical exponent $\nu=0.666...$\cite{XY}. 
Our value of $\nu$ obtained above is consistent with this value.
In fact, the same result was previously obtained in Ref.\cite{kragset},
which studied the $N=2$ model on the specific line $c_1=c_2$ in the
$c_2-c_1$ plane by means of the MC simulations of
large system sizes\footnote{The $N=2$ model with a noncompact gauge
action was studied by Motrunich et al.\cite{noncompact}.
Phase stucture of that mode was also clarified by the paper
by Kragset et al. \cite{kragset}.}.
 
\begin{figure}[thbp]
\begin{center}
\epsfxsize=7cm
\epsffile{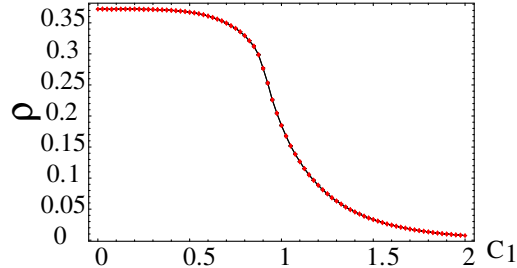}
\caption{Instanton density $\rho$ for $N=2$ 
at $c_2=0.4$ as a function of $c_1$. 
System size $L=16$.
}
\label{fig2}
\end{center}
\end{figure}

It is instructive to see the behavior of the instanton density 
$\rho$ in order to study the gauge dynamics at the phase transition
point. 
We employ the definition of $\rho$ 
in the 3D U(1) compact lattice gauge theory given by
DeGrand and Toussaint\cite{instanton}.
$\rho$ in Fig.\ref{fig2}
decreases very rapidly near the phase transition 
point at $c_1 \simeq 0.91$.
This indicates that a ``crossover" from dense to dilute instanton 
``phases" is accompanied with the phase transition. 
In other words, the observed phase transition can be interpreted 
as a confinement(small $c_1$)-Higgs(large $c_1$) phase transition.

The above conclusion is supported by the following consideration.
An effective gauge model $S_{\rm eff}(U)$
is obtained by integrating out the Higgs fields
$\phi_{xa}$ in $Z$ of Eq.(\ref{z}),
\begin{eqnarray}
Z&=&\int [dU]\exp[S_{\rm eff}(U)],\nonumber\\
\exp[S_{\rm eff}(U)]&\equiv&\int [d\phi]\exp[S(U,\phi)].
\label{Seff}
\end{eqnarray}
For small $c_1$, the above integration over the Higgs fields can be
performed by using the hopping expansion in powers of $c_1$.
The resultant effective action $S_{\rm eff}(U)$ contains {\em nonlocal} 
interaction terms of the gauge field $U_{x\mu}$.
Recently we studied models of U(1) gauge field in 3D, which contain
nonlocal interactions \cite{gauge2}.
We found that the nonlocal terms give dominant effect on the gauge dynamics
and a confinement-deconfinement phase transition takes place as their 
coefficients are getting large.

Knowledge of the phase structure of the $CP^{N-1}$ model is also useful
to identify the phases in the present model.
As explained in the introduction, the present model is the easy-plane
limit of the $CP^1$ model.
The $CP^{N-1}$ model in 3D was studied both analytically by means of
the $1/N$ expansion \cite{CP_1} and numerically by defining the model 
on the lattice \cite{CP_2,gauge3}.
These studies show that the spontaneous breaking of the internal $SU(N)$
symmetry accompanys the phase transition to the Higgs phase of the gauge 
dynamics.
Similarly in the present model, the observed second-order phase transition
corresponds to the transition from the global U(1) symmetric phase
(confinement phase) to the phase of the the spontaneous breaking of 
the global U(1) symmetry (Higgs phase).

\begin{figure}[htbp]
\begin{center}
\epsfxsize=7cm
\epsffile{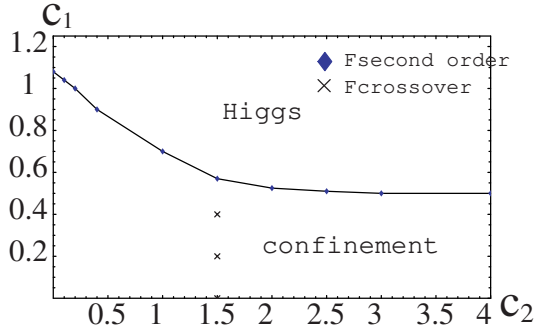}
\caption{Phase diagram for $N=2$.
There are two phases, confinement and Higgs, separated 
by a second-order phase transition line. 
There also exists 
a crossover line in the confinement phase separating dense 
and dilute instanton-density regions.
Location of the phase transition line is determined from the data of
$L=24$.
}
\label{fig3}
\end{center}
\end{figure}

In Fig.\ref{fig3}, we present 
the phase diagram for $N=2$ in the $c_2$-$c_1$ plane.
There exists a second-order phase transition line separating 
the confinement and the Higgs phases.
There also exists a crossover line similar to that in the 
3D $N=1$ U(1) Higgs model\cite{janke}.

\section{Three-flavor Higgs model ($N=3$)}
\setcounter{equation}{0}

\subsection{Symmetric case $c_{1a}=c_1$}

Let us turn to the $N=3$ case. 
Among many possibilities of three $c_{1a}$'s, 
we first consider the symmetric case $c_{11}=c_{12}=c_{13}
\equiv c_1$.
One may expect that there are two ($N-1=2$) second-order 
transitions that may coincide at a certain critical point.
Studying  the $N=3$ case is interesting from a general 
viewpoint of the critical phenomena, i.e., whether 
coincidence of multiple phase transitions changes the 
order of the transition. 
We studied various points in the $c_2-c_1$
plane and found that the order of transition 
changes as $c_2$ varies.

\begin{figure}[htbp]
\begin{center}
\epsfxsize=10.5cm
\epsffile{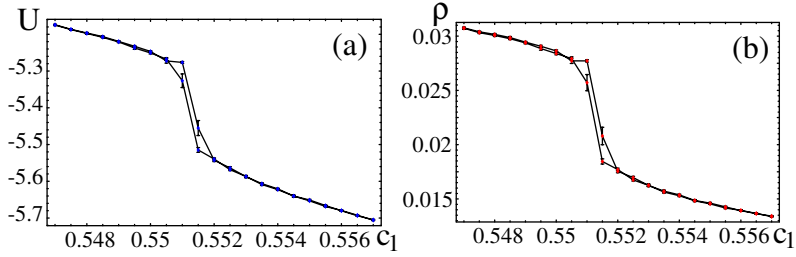}
\caption{(a) Internal energy $U$ and (b) instanton density $\rho$ 
for $N=3$ at $c_2=1.5$ and $L=16$.
Both exhibit hysteresis loops in the path
where $c_1$ is first increased and then
decreased by the step $\Delta c_1 = 0.0005$, indicating a
first-order phase transition at $c_1 \simeq 0.551$.
}
\label{fig4}
\end{center}
\end{figure}

\begin{figure}[htbp]
\begin{center}
\epsfxsize=10.5cm
\epsffile{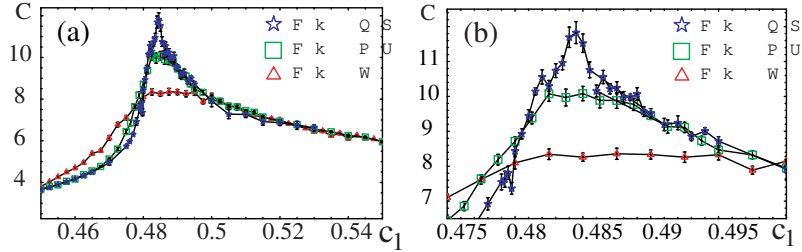}
\caption{(a) Specific heat for $N=3$ at $c_2=3.0$. 
(b) Close-up view near the peak. The peak
develops as $L$ increases.
}
\label{fig5}
\end{center}
\end{figure}

In Fig.\ref{fig4}, we show $U$ and $\rho$ along $c_2=1.5$
as a function of $c_1$.
Both quantities show hysteresis loops, which are signals of a
first-order phase transition.
In Fig.\ref{fig5}, we present $C$
at $c_2=3.0$. The peak of $C$ at around $c_1\sim 0.48$
develops as $L$ is increased, whereas $U$ shows no 
discontinuity and hysteresis.
Therefore, we conclude that the phase transition at 
$(c_2,c_1)\sim (3.0, 0.48)$ is of second order.

\begin{figure}[htbp]
\begin{center}
\epsfxsize=6.5cm
\epsffile{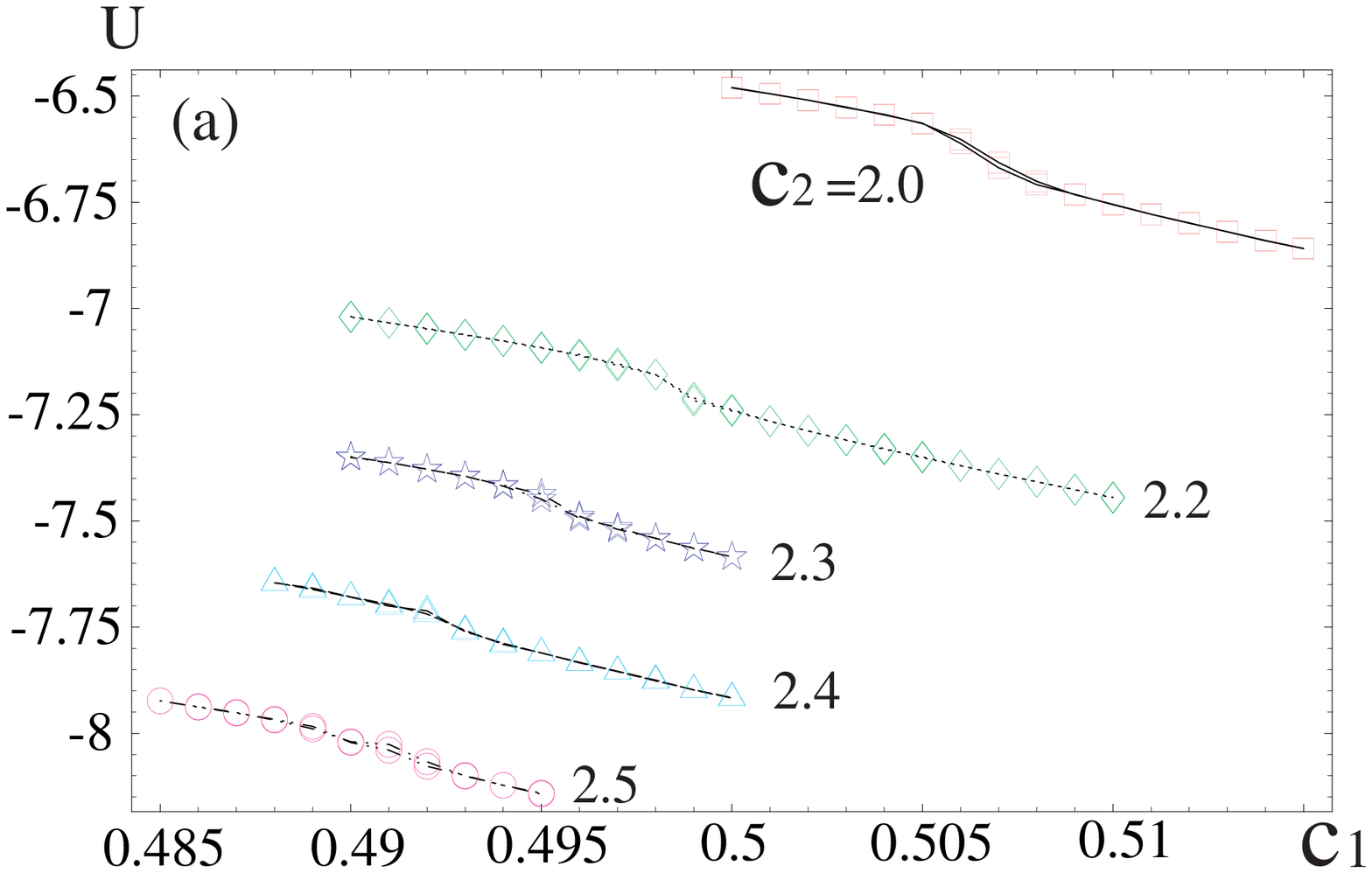}
\epsfxsize=6cm
\epsffile{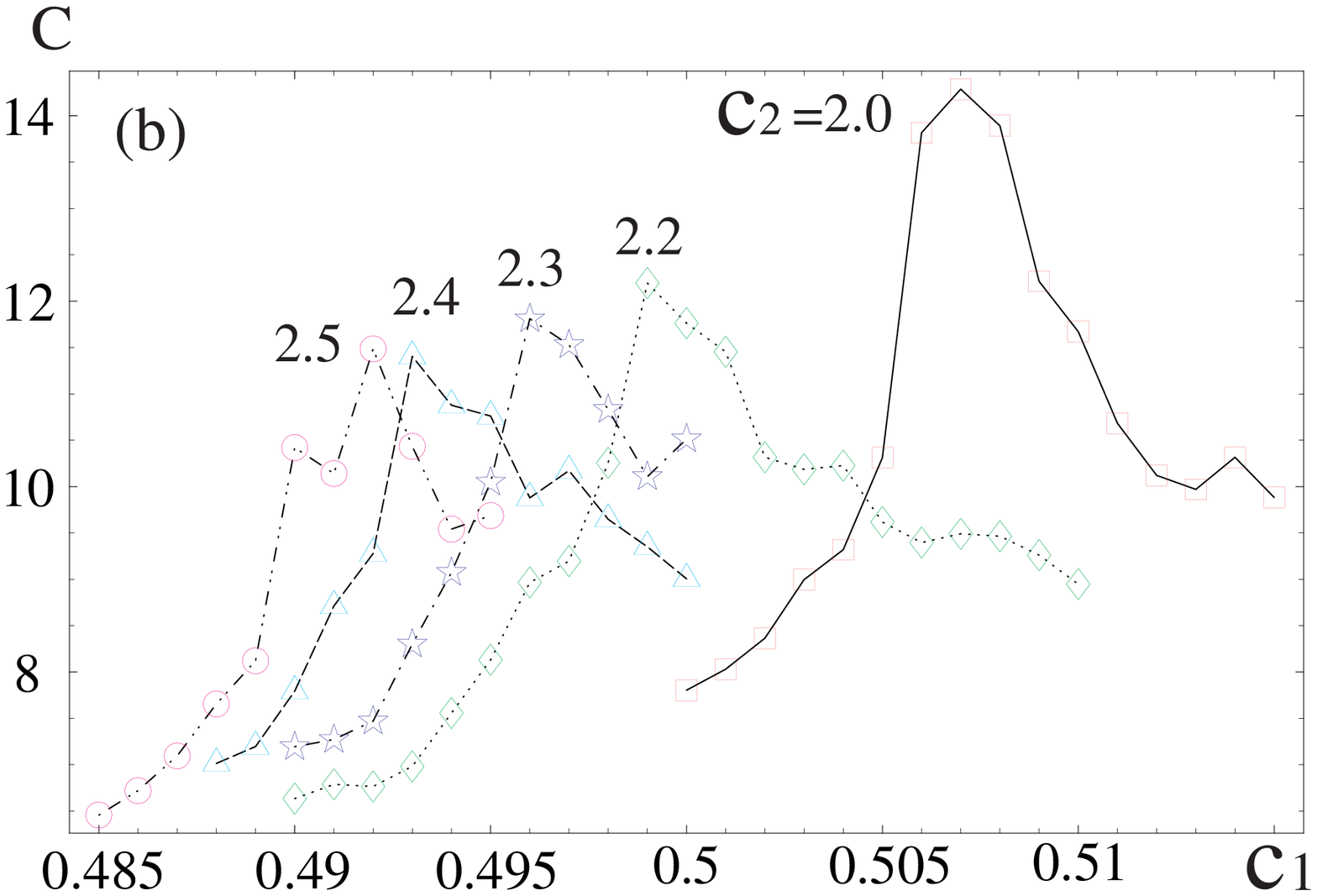}
\caption{(a) Internal Energy $U$ and (b) Specific heat $C$
for $N=3$ and $L=32$ at $c_2=2.0\sim 2.5$. (Values of $c_2$ are indicated
near each curve.)
As $c_2$ increases, the changes of $U$  become
milder and the region  and size of the peak of $C$ become reduced.}
\label{fig6}
\end{center}
\end{figure}

\begin{figure}[htbp]
\begin{center}
\epsfxsize=6.5cm
\epsffile{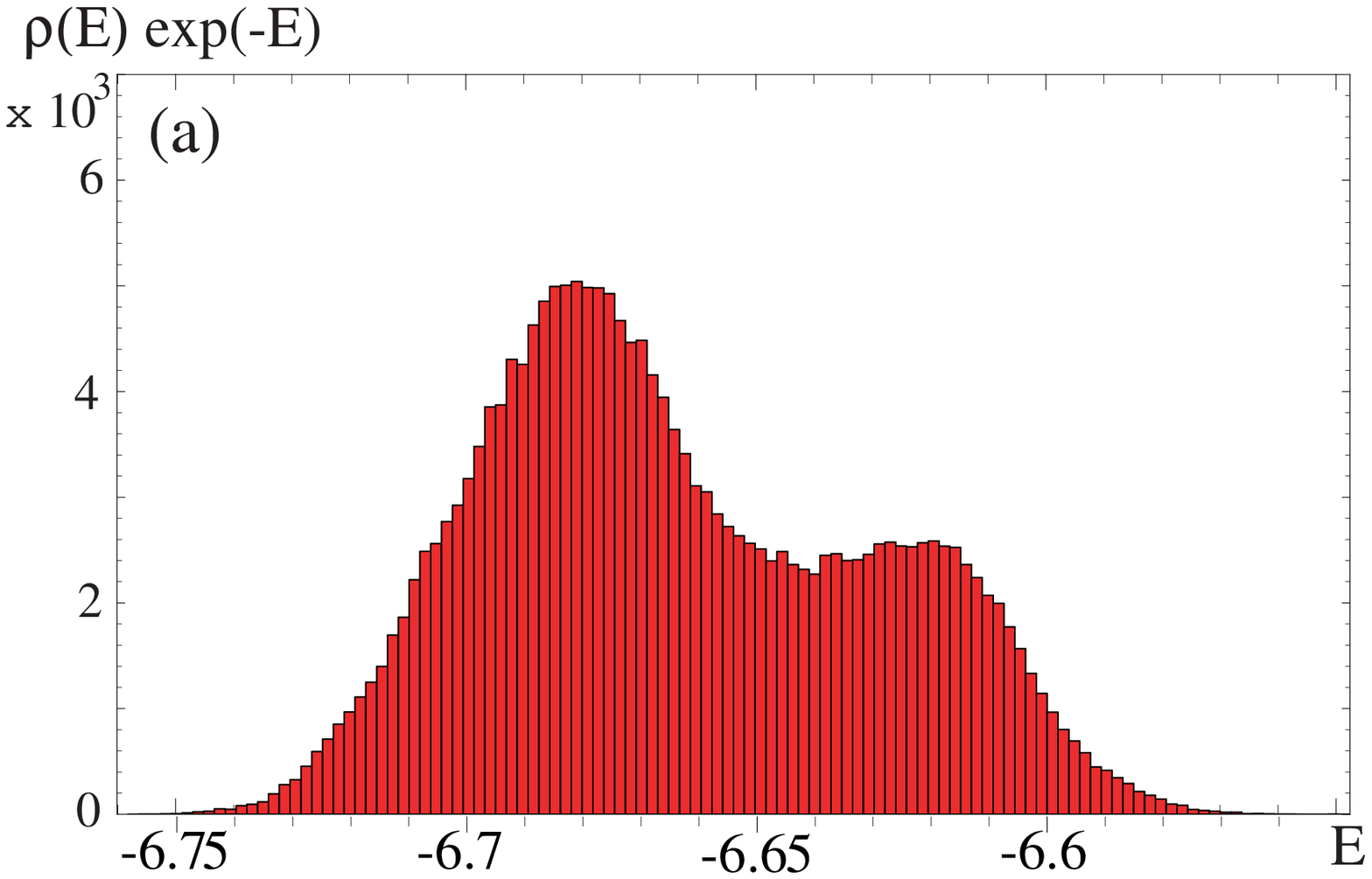}
\epsfxsize=6.5cm
\epsffile{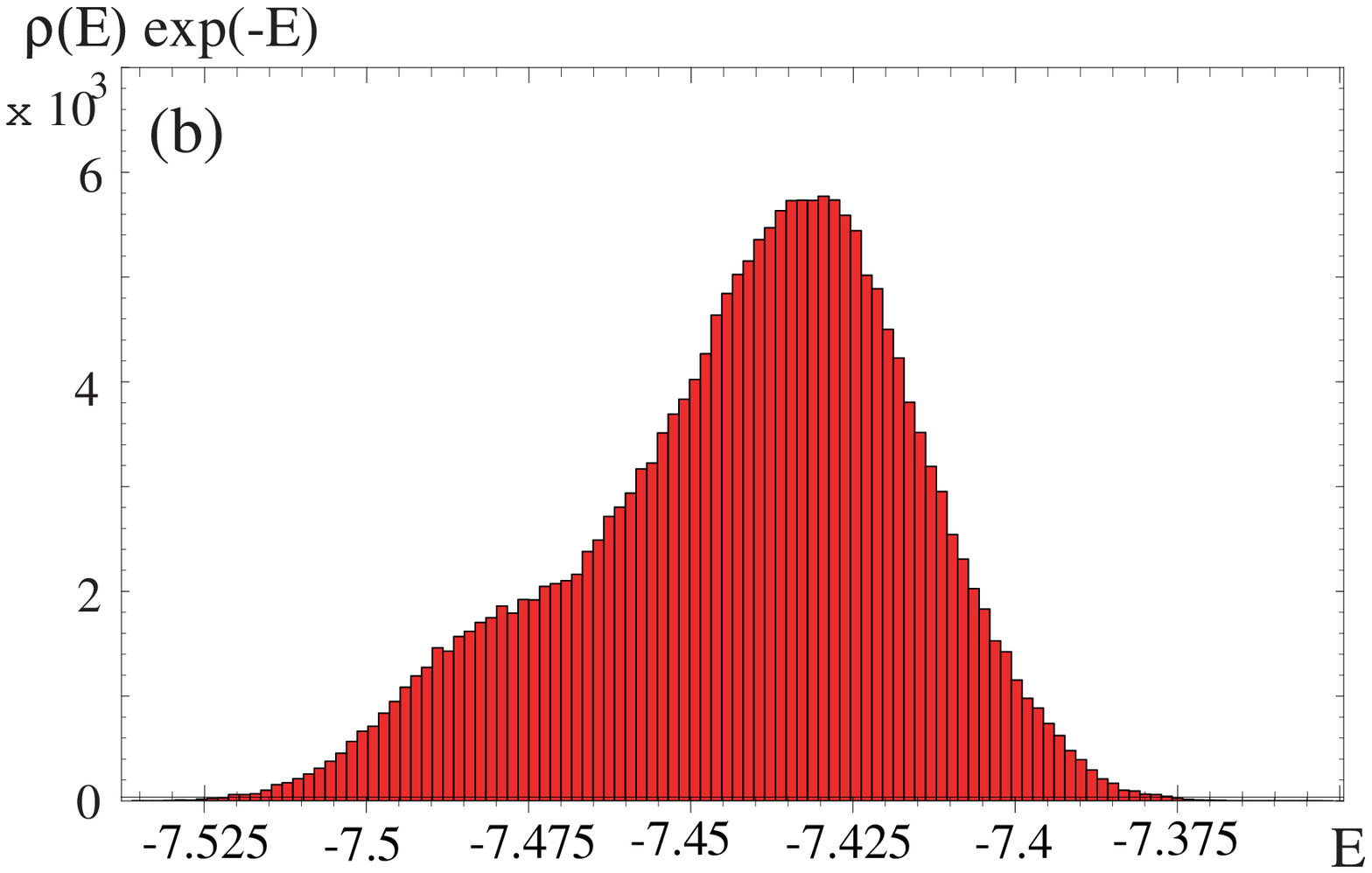}\\
\epsfxsize=6.5cm
\epsffile{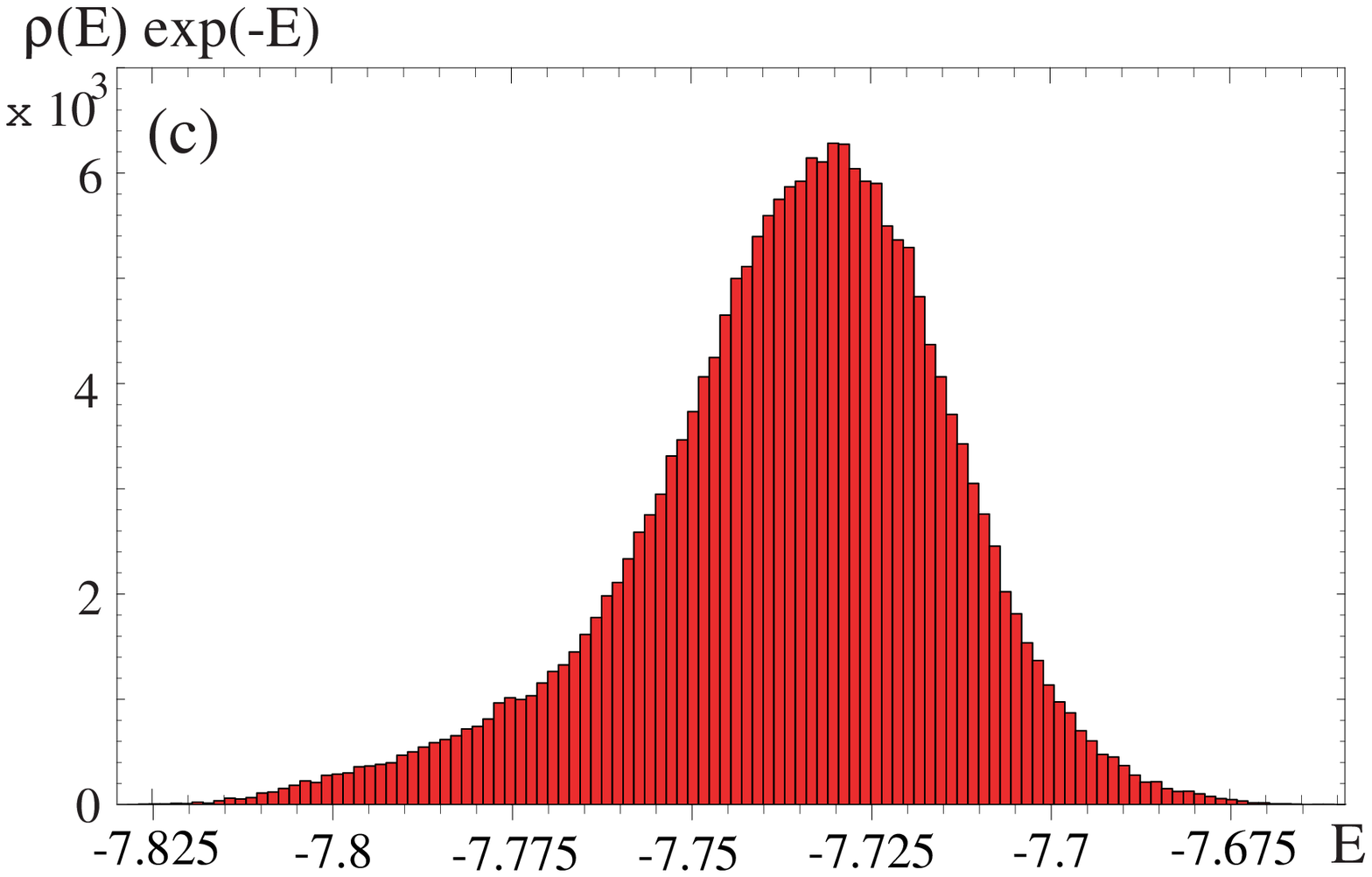}
\epsfxsize=6.5cm
\epsffile{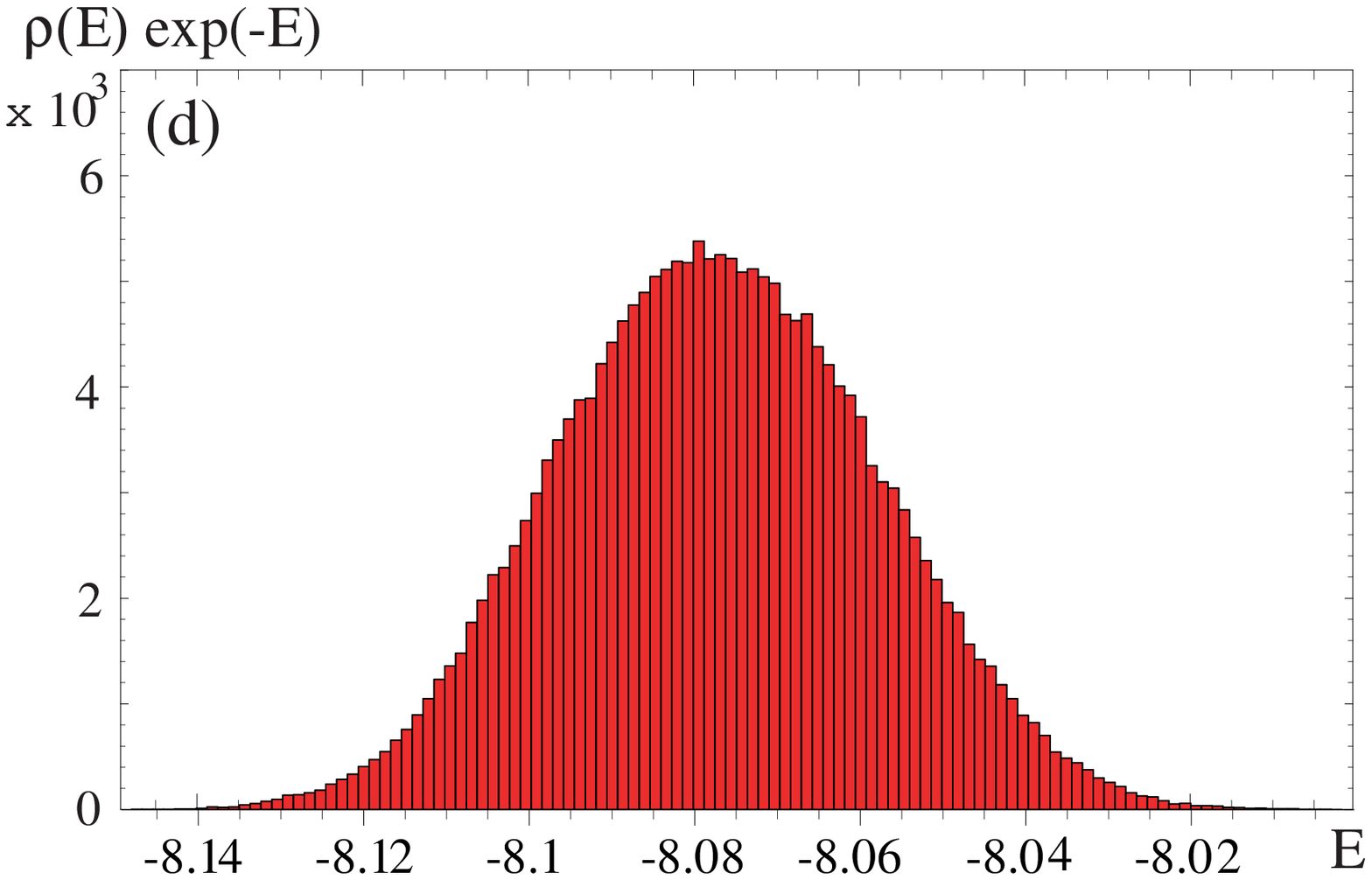}\\
\caption{Distribution of the energy, $\rho(E) \exp(-E)$ of 
Eq.(\ref{rho(E)}), for $N=3$ and $L=32$. 
(a)  $(c_2,c_1)=(2.0, 0.507)$;
(b)  $(c_2,c_1)=(2.3, 0.495)$;
(c)  $(c_2,c_1)=(2.4, 0.493)$;
(d)  $(c_2,c_1)=(2.5, 0.492)$.
The signal of double-peak structure, which is shown clearly
in Fig.(a),  becomes weaker as $c_2$ increases, and
disappears in Fig.(d).}
\label{fig7}
\end{center}
\end{figure}

In order to locate the tricritical point from the first to
second phase transitions, we studied the region $c_2 = 2.0 \sim 2.5$
in detail.
In Fig.\ref{fig6}, we present
the internal energy $U$ and the specific heat $C$ at
$c_2=2.0, 2.2, 2.3, 2.4, 2.5$ 
for the system size $L=32$.
$U$ at $c_2=2.0$ shows a hysteresis at $c_1 \simeq 0.506 \sim 0.508$.
As $c_2$ increases, the hysteresis becomes milder and
at $c_2=2.5$ it almost disppears.
Also, as $c_2$ increases,  the region and height of the peak of $C$ measured 
from the smooth background become reduced.
These behavior, together  with the size dependence of 
$U$ and $C$,  suggest that the change in the  order of phase transition
from the first-order one to the second-order one
takes place in this region of $c_2$.
To support this point,  
we also studied the distribution  $\rho(E)\exp (-E)$ of the 
internal-energy, which is defined as 
\begin{equation}
Z=\int [dU][d\phi]\exp (S)
=\int dE \int [dU][d\phi]\exp (S)\delta(S+E)
=\int dE \; \rho(E)\exp (-E).
\label{rho(E)}
\end{equation}
\noindent
If the phase transition is 
first order, $\rho(E)\exp (-E)$  exhibits a double peak
structure near the critical point, 
while a second-order transition exhibits a single peak structure. 
In Fig.\ref{fig7}, we present the distribution
of $\rho(E)\exp (-E)$  
for $c_2=2.0\sim 2.5$ and $L=32$.

\newpage
\noindent
The value of $c_1$ is chosen near the peak location of $C$
given in Fig.\ref{fig6}(b).
Apparently, Fig.\ref{fig7}(a), the distribution at $c_2= 2.0$,
fits better by a double-peak (e.g., double Gaussian) 
distribution rather than by a single-peak one.
The fact that two peaks here have different weights mainly  reflects that
it is slightly away from the critical point of $c_1$. 
Fig.\ref{fig7}(c) for $c_2=2.4$
has a single peak at $E \simeq -7.73$, but 
not symmetric around this peak, which shows a remnant 
of the second peak at lower $U$ region.
Fig.\ref{fig7}(d) for $c_2=2.5$ shows a clean
single-peak distribution.
From these observations we  
determine that 
the tricritical point is located at $c_2=c_{2{\rm tc}}\simeq 2.4\pm 0.1$.

\begin{figure}[htbp]
\begin{center}
\epsfxsize=11cm
\epsffile{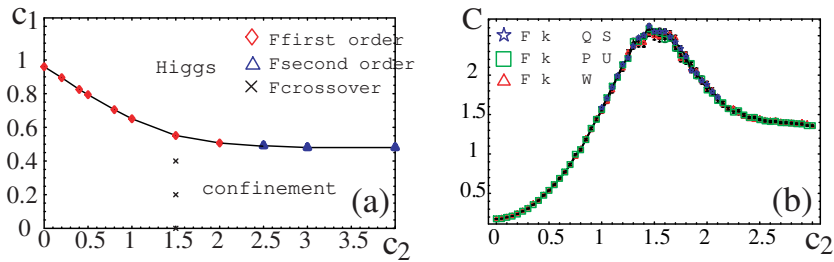}
\caption{(a) Phase diagram for the $N=3$ symmetric case.
The phase transitions are first order in the
 region  $c_2 < c_{2{\rm tc}}\simeq 2.4$, 
 whereas they are second order in the region 
 $c_2 > c_{2{\rm tc}}$. 
There exists a 
tricritical point at around $(c_2,c_1)\sim (2.4,0.49)$.
Crosses near $c_2 = 1.5$ line show crossovers.
(b) Specific heat for $N=3$ at $c_1=0.2$.
It has a system-size independent smooth peak at which
a crossover takes place.
}
\label{fig8}
\end{center}
\end{figure}

\begin{figure}[htbp]
\begin{center}
\epsfxsize=6cm
\epsffile{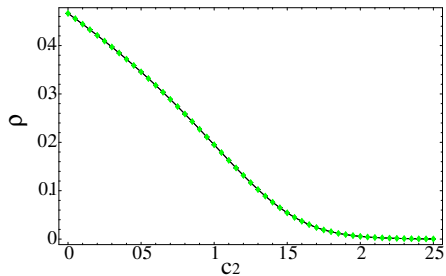}
\caption{Instanton density $\rho$ in the
$N=3$ symmetric case for $c_1=0.2$ 
as a function of $c_2$. System size $L=16$.
}
\label{fig9}
\end{center}
\end{figure}

In Fig.\ref{fig8}(a), we present the phase diagram of 
the symmetric case for $N=3$, where the order 
of transition between the confinement and Higgs phases changes
from first ($c_2 < c_{2{\rm tc}}$) to second order ($c_{2{\rm tc}} < c_2$).  
In Fig.\ref{fig8}(b) we present $C$ along $c_1 = 0.2$, which
shows a smooth nondeveloping peak. 
As shown in Fig.\ref{fig9}, the instanton density $\rho$ decreases 
smoothly around this peak.
These results indicate a crossover at $c_2 \simeq 1.5$.

\subsection{Asymmetric case}

Then it becomes interesting to consider 
asymmetric cases, e.g., $c_{11}\neq c_{12}=c_{13}$.
This case is closely related to a doped AF magnet. 
$\phi_2$ and $\phi_3$ correspond there to the $CP^1$ spinon
field in the deep easy-plane limit, whereas $\phi_1$ corresponds 
to doped holes (although they are fermionic). 
This case is also relevant to cosmology because 
the order of Higgs phase transition in the early universe is 
important in the inflational cosmology.
Furthermore, one may naively expect that once a 
phase transition to the Higgs phase occurs at certain 
temperature $T$, 
no further phase transitions take place at lower $T$'s 
even if the gauge field couples with other Higgs bosons.
However, our investigation below 
will show that this is not the case.

\begin{figure}[htbp]
\begin{center}
\epsfxsize=12cm
\epsffile{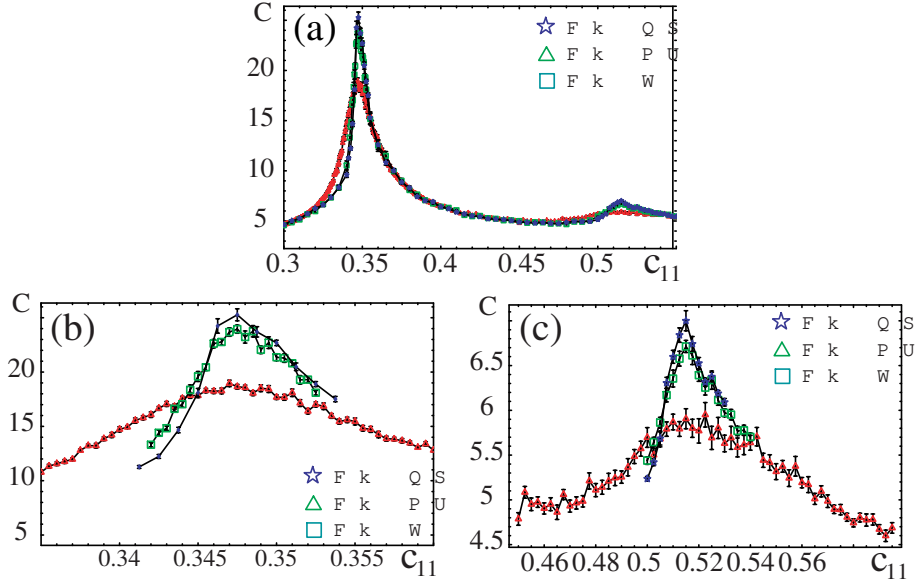}
\caption{(a) Specific heat of the $c_1=(1,2,2)$ model ($N$=3) at $c_2=1.0$.
(b,c) Close-up views of $C$ near (b) $c_{11} \sim 0.35$ 
and (c) $c_{11}\sim 0.52$.
}
\label{fig10}
\end{center}
\end{figure}

Below we shall consider the two cases, 
(i) $c_{12}=c_{13} > c_{11}$
and (ii) $c_{12}=c_{13} < c_{11}$.
Let us first consider the case (i) $c_{12}=c_{13}=2c_{11}$,
which we call the $c_1=(1,2,2)$ model, and focus on the case
$c_2=1.0$. 
As shown in Fig.\ref{fig10}(a), 
$C$ exhibits two peaks at $c_{11}\sim 0.35$ and $0.52$.
Figs.\ref{fig10}(b),(c) present the detailed behavior 
of $C$ near these peaks, which show 
that the both peaks develop as $L$ is increased.
We conclude that both of these peaks show second-order
transitions. 
This result is interpreted as 
the first-order phase transition 
in the symmetric 
$N=3$ model is decomposed into two second-order transitions
in the $c_1=(1,2,2)$ model.

\begin{figure}[h]
\begin{center}
\epsfxsize=6.5cm
\epsffile{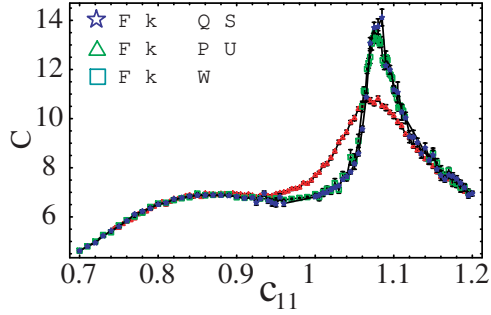}
\caption{Specific heat of the $c_1=(2,1,1)$ model at $c_2=1.0$.
}
\label{fig11}
\end{center}
\end{figure}

Let us turn to the opposite case (ii), 
$c_{12}=c_{13}=0.5c_{11}$, 
i.e., the $c_1=(2,1,1)$ model at $c_2=1.0$.
One may expect that two second-order phase transitions 
appear as in the previous $c_1=(1,2,2)$ model.
However, the result shown in Fig.\ref{fig11} indicates 
that there exists only one second-order phase transition near 
$c_{11}\sim 1.08$.
The broad and smooth peak near $c_{11}\sim 0.85$ shows no $L$ 
dependence and we conclude that it is a crossover.
This crossover is similar to that 
in the ordinary $N=1$ gauge-Higgs system 
as we shall see by the measurement of $\rho$ below.

The orders of these transitions are understood as follows: 
In the $c_1=(1,2,2)$ model, as we increase $c_{11}$,
the two modes $\phi_{xa} (a=2,3)$ with larger $c_{1a}$ firstly 
become relevant 
and the model is effectively the symmetric $N=2$ model.
The peak in Fig.\ref{fig10}(b) is interpreted as that of the 
second-order phase transition in this model. 
As the Higgs couplings $c_{11}$'s are increased further, 
the fluctuations of the gauge field is negligibly small, 
and the effective model is the $N=1$  
$XY$ model of  $\phi_{x1}$. It  gives 
 the second-order peak in Fig.\ref{fig10}(c).
Similarly, in the $c_1=(2,1,1)$ model,  $\phi_{x1}$ firstly becomes 
relevant.
The effective model is the   $N=1$ model, which gives 
the broad peak in the specific heat $C$ in Fig.\ref{fig11} as 
the crossover takes place there\cite{janke}. For larger values of $c_{11}$'s,
the effective model is the $N=2$ symmetric model of $\phi_{x2},
\phi_{x3}$ and $U_{x\mu}$, giving  the sharp second-order 
peak in $C$ in Fig.\ref{fig11}.

\begin{figure}[htbp]
\begin{center}
\epsfxsize=11cm
\epsffile{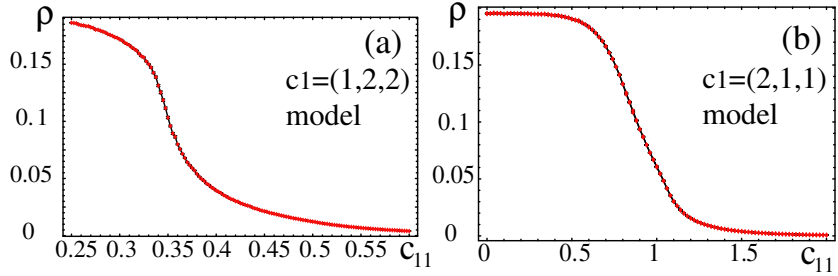}
\caption{Instanton density $\rho$ at $c_2=1.0$ in the (a) $c_1=(1,2,2)$ 
model and (b) $c_1=(2,1,1)$ model. System size $L=16$.
}
\label{fig12}
\end{center}
\end{figure}

In Fig.\ref{fig12}, we present $\rho$
of the $c_1=(1,2,2)$ and $(2,1,1)$ models
as a function of $c_{11}$.  
$\rho$ of the $c_1=(1,2,2)$ model  
decreases very rapidly at around $c_{11}\sim 0.35$, 
which is the phase transition point in lower $c_{11}$ region.
On the other hand,  at the higher phase
transition point, $c_{11}\sim 0.52$, $\rho$ shows
no significant changes.
This observation indicates that the lower-$c_{11}$ 
phase transition is the confinement-Higgs transition, whereas 
the higher-$c_{11}$ transition is a charge-neutral 
$XY$-type phase transition.

On the other hand, $\rho$ of the $c_1=(2,1,1)$ model 
decreases rapidly at around $c_{11}\sim 0.85$,
where $C$ exhibits a broad peak.
This indicates that the crossover from the dense to 
dilute-instanton regions occurs there just like in 
the $N=1$ case\cite{janke}.
No ``anomalous" behavior of $\rho$ is observed at the critical 
point $c_{11} \sim 1.1$, and therefore the phase transition 
is that of the neutral mode. 

\setcounter{equation}{0}
\section{Symmetric model with $N=4$ and $5$ }

\begin{figure}[hbtp]
\begin{center}
\epsfxsize=6cm
\epsffile{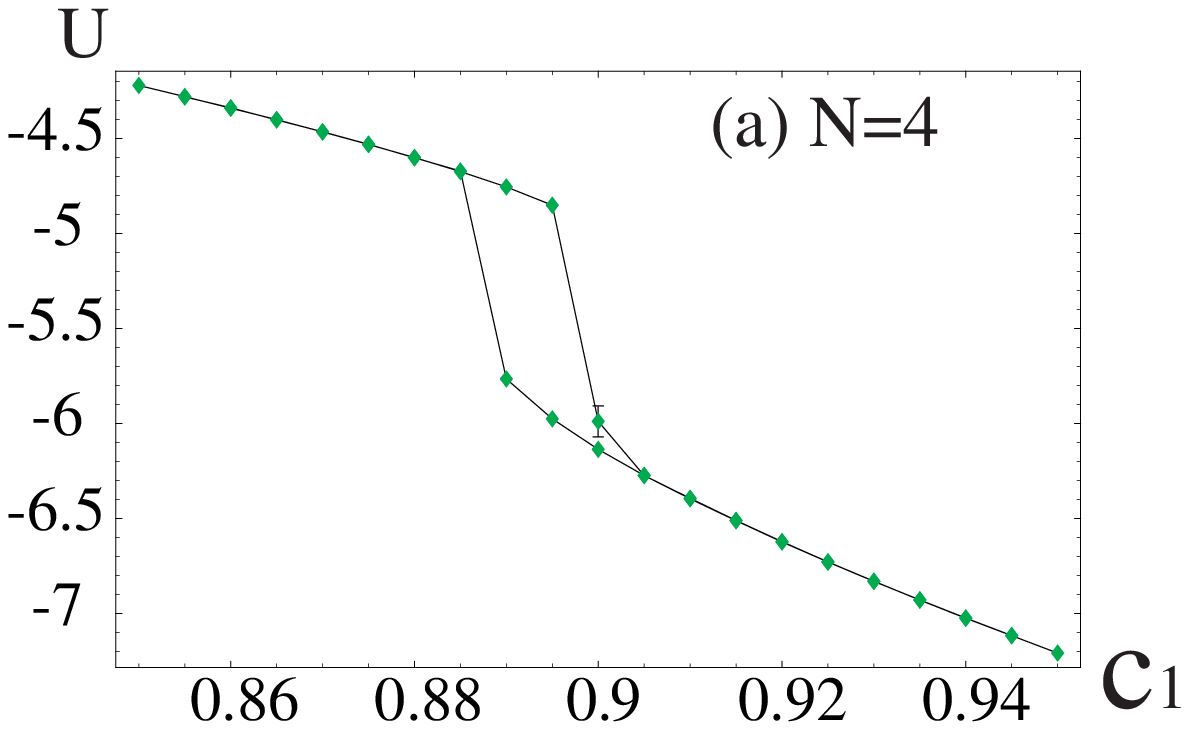}
\hspace{1cm}
\epsfxsize=6cm
\epsffile{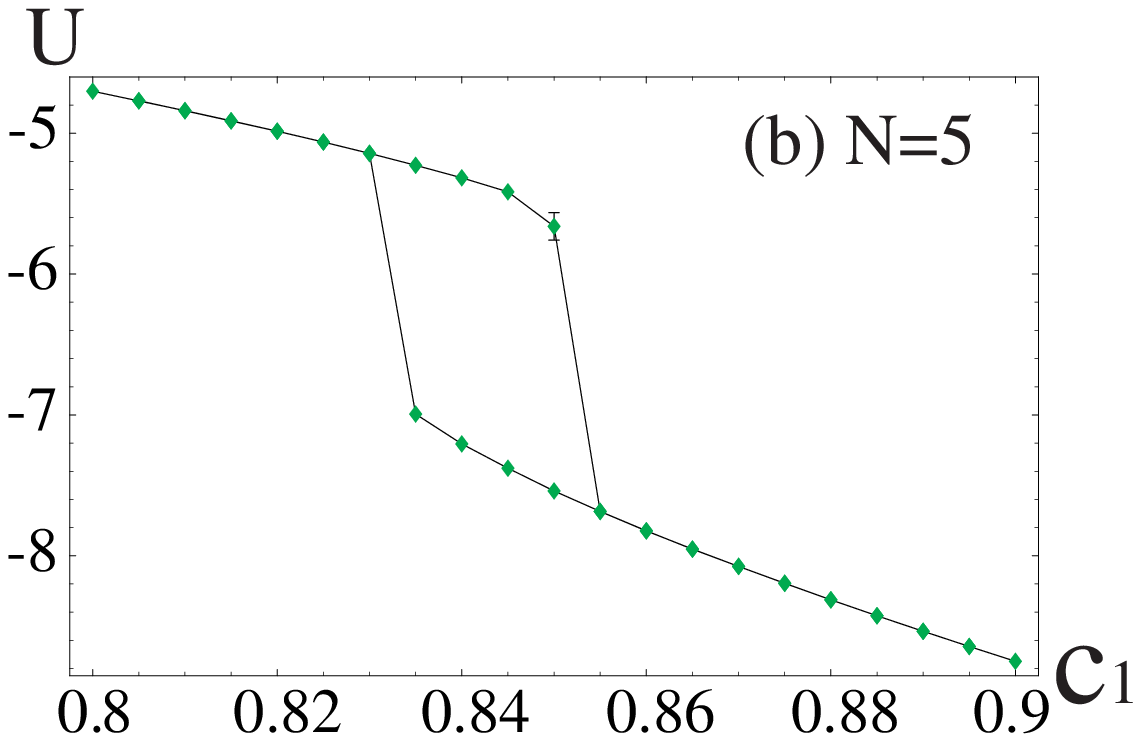}
\vspace{-0.5cm}
\caption{Internal energy $U$ in (a) the  $N=4$ model and
(b) $N=5 $ model for $c_2=0$
with the system size $L=24$.
Both exhibit hysteresis loops in the path
where $c_1$ is first increased and then
decreased by the step $\Delta c_1 = 0.005$.
This indicates  a first-order phase transition for $c_2=0$
at $c_1 \simeq 0.89$ for $N=4$  and at $c_1 \simeq 0.84$
for $N=5$.
}
\label{fig13}
\end{center}
\end{figure}

We have also studied the symmetric case for
$N=4$ and $5$ multi-Higgs models at $c_2=0$.
Both cases show clear signals of 
first-order transitions at $c_1 \simeq 0.89 (N=4),
0.84 (N=5)$ as shown in Fig.\ref{fig13}. 
On the other hand, at  $c_2=\infty$,
the gauge dynamics is ``frozen" to $U_{x\mu}=1$
up to gauge transformations, so 
there remain $N$-fold {\em independent} $XY$ spin models,
each of which exhibits a second-order transition at 
$c_1 \simeq 0.46$. Thus we expect a tricritical point 
for general $N > 2$ at some {\it finite} $c_2$
separating first-order and second-order transitions. 

\section{Conclusion}
\setcounter{equation}{0}

In the present paper, we studied the U(1) multi-flavor Higgs model
in 3D, which is closely related to various interesting physical 
systems. 
By means of the MC simulations,
we clarified its phase structure and critical behavior.
Let us summarize the results.
For $N=2$ there is a critical line $\tilde{c}_{1}(c_2)$ of  
second-order transitions in the $c_2-c_1$ plane,
which distinguishes the Higgs phase ($c_1 > \tilde{c}_{1}$)
and the confinement phase ($c_1 < \tilde{c}_{1}$).
This result is consistent with Kragset et al.\cite{kragset}.
We obtained the crtitical exponent of the phase transition
by means of the FSS and found that the result is very close to 
that of the 3D XY model.

For $N=3$ there is a similar transition line, but the region
$0 < c_2 < c_{2{\rm tc}} \simeq 2.4$ is of {\em first-order}
transitions while the region $c_{2{\rm tc}} < c_2$ is
of {\it second-order}  transitions. 
We concluded that there exists a tricritical point.
It is very interesting and also important to clarify 
the nature of the tricritical point, especially its critical 
exponent.
This problem is under study and the result will be published
in future .

To study the mechanism of generation of these first-order 
transitions, we studied the asymmetric cases and found 
two second-order transitions [in the 
$c_1=(1,2,2)$ model] or one crossover
and one second-order phase transition 
[in the $c_1=(2,1,1)$ model]. 
The former case implies that two simultaneous second-order 
transitions strengthen the order to generate a first-order 
transition.
Chernodub et al.\cite{chernodub} reported
a similar generation of an enhanced first-order transition
in a related 3D Higgs model 
with singly and doubly charged scalar fields. 
We stress that the above  change of the order is dynamical
because (1) It depends on the value of $c_2$,
(2) Related 3D models, the $CP^{N-1}$ and $N$-flavor
$CP^1$ gauge models, exhibit always second-order 
transitions (See Ref.\cite{gauge3}).

\begin{center}
{\bf Acknowledgements}
\end{center}
We thank K. Sakakibara and M.N.Chernodub for useful discussions.
We also thank Y. Nakano for his assistance in numerical calculations.
This work was partially supported by Grant-in-Aid
for Scientific Research from Japan Society for the 
Promotion of Science under Grant No.20540264.


\end{document}